%% file: main.tex
\documentclass[10pt,conference]{IEEEtran}
\IEEEoverridecommandlockouts
\usepackage{cite}
\usepackage{amsmath,amssymb,amsfonts}
\newtheorem{definition}{Definition}
\usepackage{algorithmic}
\usepackage{graphicx}
\usepackage{textcomp}
\usepackage{xcolor}

\usepackage[hyphens]{url}
\usepackage{breakurl}

\usepackage{tcolorbox}

\newcommand{\deltaup}[1]{\textcolor{green!50!black}{$\uparrow$#1}}
\newcommand{\deltadown}[1]{\textcolor{red}{$\downarrow$#1}}

\makeatletter
\newcommand{\linebreakand}{%
  \end{@IEEEauthorhalign}
  \hfill\mbox{}\par
  \mbox{}\hfill\begin{@IEEEauthorhalign}
}

\makeatother

\usepackage{amssymb}
\usepackage{booktabs}
\usepackage{tabularx}
\usepackage{comment}
\usepackage{enumitem} 
\usepackage{multirow}
\usepackage{graphicx}
\def\BibTeX{{\rm B\kern-.05em{\sc i\kern-.025em b}\kern-.08em
    T\kern-.1667em\lower.7ex\hbox{E}\kern-.125emX}}
\begin{document}



\title{How effective are traditional test criteria at detecting bugs in large language models generated code?}

\author{\IEEEauthorblockN{Asma Hamidi}
\IEEEauthorblockA{\textit{SnT, University of Luxembourg} \\
Luxembourg \\
asma.hamidi@uni.lu}
\and
\IEEEauthorblockN{Michael Konstantinou}
\IEEEauthorblockA{\textit{SnT, University of Luxembourg} \\
Luxembourg \\
michael.konstantinou@uni.lu}
\and
\IEEEauthorblockN{Renzo Degiovanni}
\IEEEauthorblockA{\textit{Luxembourg Institute of Science} \\
\textit{and Technology} \\
Luxembourg \\
renzo.degiovanni@list.lu}

\linebreakand
\IEEEauthorblockN{Mike Papadakis}
\IEEEauthorblockA{\textit{SnT, University of Luxembourg} \\
Luxembourg \\
michail.papadakis@uni.lu}
}

\maketitle

\begin{abstract}


Test adequacy criteria are widely used to evaluate and guide software testing. Although prior research has extensively examined these criteria using human-written programs, faults, and tests, the increasing adoption of Large Language Models (LLMs) for code generation raises important questions about their effectiveness in detecting LLM-induced faults. To investigate this, we conduct an empirical study involving 5 LLMs and 4 benchmarks, simulating end-to-end workflows in which both code and tests are automatically generated. We collect  6,000+ faulty program instances and evaluate the effectiveness and efficiency of 3 widely used adequacy criteria: statement coverage, branch coverage, and mutation testing. Our findings reveal several key insights. First, most faults introduced by LLMs are relatively trivial to catch. Second, the challenging faults are difficult to trigger using either traditional coverage-based or mutation-based criteria. Third, actual fault detection rates remain extremely low, often near zero, because test oracles fail to capture faulty behavior triggered by the generated test prefixes, exposing a critical limitation of automated test generation. Fourth, prompt-aware oracles can improve fault detection, but their overall effectiveness remains limited, highlighting the need for users to manually reason about test assertions. We further observe that mutation testing only marginally outperforms traditional coverage criteria in both triggering and detecting faults, raising questions about whether its significantly higher application cost is justified in this context.
\end{abstract}


\section{Introduction}

The question of the sufficiency and effectiveness of coverage criteria in steering software testing toward fault detection in automatically generated code remains unresolved. Prior research has examined the relationship between various structural coverage measures and fault detection using programs and defects produced by human developers \cite{10.1145/125489.125473,chekam2017empirical}. Determining whether traditional test coverage adequacy criteria remain suitable for automatically generated code is increasingly important, particularly with the growing adoption of ``vibe coding''.

Traditional test adequacy criteria are designed to capture mistakes typically made by human developers \cite{ammann2017introduction}. They implicitly assume that faults arise around conditional boundaries, that code complexity is manageable and easily understandable by human readers, that code is coherent with limited redundancy, and that there is a clear separation of concerns within the program logic. However, these assumptions may not hold for automatically generated code. Even if they do, it remains unclear whether such code exhibits the same fault patterns as human-written programs. For example, errors in LLM-generated code often stem from under-specified prompts \cite{larbi2025prompts}, ambiguities and contradictions in natural language descriptions \cite{10.1007/s10664-025-10614-4}, or the model’s inability to correctly align code with its intended semantics \cite{ZhuLLZLJ024}.


The test effectiveness of traditional adequacy criteria remains under-explored and insufficiently characterized in this scenario. Should these criteria prove to be of limited significance, this would motivate new research directions aimed at developing specialized testing metrics tailored to LLM-based or agent-based systems. Moreover, such findings would raise concerns related to the scientific validity of prior studies that rely on these criteria to assess the effectiveness of test generation or oracle construction techniques. Our empirical analysis provides strong evidence that traditional test criteria fail to trigger and detect almost all faults examined (approximately miss approximately 99\% of them), suggesting that they are of limited  power. 

We start our study by building the pool of LLMs-generated faults to investigate. 
To do so, we employ four established coding benchmarks (namely \textit{HumanEval+}~\cite{liu2024your,chen2021evaluating}, \textit{MBPP}~\cite{austin2021program}, \textit{BigCodeBench}~\cite{zhuo2025bigcodebench} and \textit{NaturalCodeBench}~\cite{zhang2024naturalcodebench}), for which we augment the provided test suites with LLM-generated tests. These augmented test suites are generated with the intention achieve near-adequacy with respect to standard criteria, whose coverage and mutation scores achieved ensure a high-quality pool of test data. 
We then run the code generation process using different LLMs (namely, GPT-5-Mini, GPT-4.1-Mini, Claude-Haiku 4.5, Deepseek-v4-flash, and Llama 3.3-70B-Instruct) accordingly. 
We then use the augmented test pool to identify faulty cases by comparing the behavior of the generated programs against that of a reference (assumed correct) implementation. This controlled setup ensures that any fault triggered by the tests can be reliably detected, while avoiding dependence on the specific test oracles (i.e., assertions). 
Overall, we collect a total of 73,785 valid faulty LLMs-generated implementations, from which we remove 43,272 faults that are relatively trivial to detect (i.e., they can be revealed by almost any test) and can bias our evaluation. 
Hence, we end up with a substantial subset of 30,513 challenging faults. However, to avoid having an imbalanced number of faults for different tasks, we only consider the most difficult fault per task, resulting in a final set of 6,066 more challenging faults that can potentially escape standard testing processes.

To study  whether traditional test adequacy criteria remain suitable for automatically generated code, we perform the following controlled experiment. 
For each test adequacy criterion under study (statement/branch/mutation), we randomly sample 100 (minimal) test suites from the generated test pool that maximize the coverage metric of interest (i.e. we interactively pick a random test that increases the coverage, until same coverage as the generated test pool is reached). 
We then evaluate the sampled test suites in terms of both \emph{fault triggering} and \emph{fault detection} capabilities. We distinguish between these two notions: fault triggering assesses whether a test input exercises a faulty execution path, whereas fault detection evaluates whether the associated test oracle successfully identifies the resulting erroneous behavior. This distinction is critical, as incorrect or insufficient test oracles may fail to detect faults even when they are triggered.

Perhaps the most notable finding of our study is the relatively low fault-triggering capability of the considered criteria, as well as the marginal differences between them. 
On average, the fault triggering rate for statement coverage, branch coverage, and mutation testing is 32.9\%, 38.9\%, and 32.2\%, respectively, while their respective fault detection rate is approximately 1-2\%. 
To further strengthen fault detection, we designed a prompt-aware oracle generation approach, following the suggestions made by prior work \cite{huang2024measuring}, by asking LLMs to augment tests and oracles based on the given task descriptions/prompts. While this approach shows some potential, the overall effectiveness remains limited (we observe a fault detection improvement of around 6.6\% on average). These findings highlight the continued need for human oversight, particularly in the construction and validation of test assertions.

Another interesting (and perhaps surprising) findings of our study is that while there is a lot of evidence that traditional fault injection techniques can produce artificial faults that couple with real (human-written) faults~\cite{10136793}, results suggest that these do not couple with faults generated by the LLMs. In a sense, LLMs make a different kind of mistakes than humans, which may require adapting and rethinking the software testing methodologies to test LLM-based and agent-based code generators.

Further research is required to deepen our understanding of this fundamental aspect of software testing, and we do not claim to have fully resolved all related questions in this work. Nevertheless, we believe that our findings substantially advance the understanding of coverage criteria in the context of ``vibe coding'' and automatically generated programs. Our primary contributions are twofold: first, we highlight a critical challenge in transitioning from traditional testing practices to settings involving machine-generated code; second, we expose the limitations of conventional adequacy criteria and outline promising directions for future research. Most importantly, our study provides empirical evidence of a lack of effective guidance toward strong fault detection in code generated by large language models.

\section{Background and Related Work}

\subsection{Test Adequacy Criteria}

In 1975, Goodenough and Gerhart~\cite{6312836} wondered what constitutes an adequate test. That is, what criterion can be used to determine the quality of a test. They argue that a test should fulfill two major requirements: \emph{reliability} (a test can be trusted if it always produces the same output) and \emph{validity} (if a test produces a meaningful result). 
Their work led us to the notion of \emph{test adequacy criteria}~\cite{10.1145/267580.267590}, defining  metrics to determine whether a software test suite is sufficiently thorough for a given testing objective. 
They define measurable conditions that a test suite is expected to satisfy. 

Popular adequacy criteria include code-based criteria (e.g., statement, branch, and path coverage), specification-based criteria (e.g., requirements or scenario coverage), and fault-based criteria such as mutation testing. 
This remains an active area of research~\cite{someoliayi2019program, molina2025state}, and numerous criteria have been proposed and evaluated over the years. In this study, we focus on the three most widely adopted criteria in both academic research and software engineering practice: \emph{statement coverage}, \emph{branch coverage}, and \emph{mutation-based adequacy criteria}.

\textbf{Statement Coverage} (or \textbf{Line Coverage}) requires that every executable statement in the program under test be executed by at least one test case. While simple and widely adopted, statement coverage is generally considered a relatively weak adequacy criterion~\cite{myers2006art}, as executing a statement does not guarantee that all relevant execution scenarios have been explored. In particular, the same statement may be exercised under different control-flow paths and input conditions, many of which may remain untested despite full statement coverage.

\textbf{Branch Coverage} (or \textbf{Decision Coverage}) is a stronger alternative~\cite{10.1145/267580.267590, myers2006art}. It requires that a test suite exercises every branch in the control-flow graph of the program under test. In practice, this means that each decision outcome (e.g., both the true and false outcomes of a conditional statement) must be executed by at least one test case.

\textbf{Mutation Testing} evaluates a test suite thoroughness by injecting small syntactic changes into the program under test to create artificial faults called mutants\cite{PAPADAKIS2019275}. Given a test suite, a mutant is killed if at least one test produces a different outcome on the mutant than on the original program, otherwise the mutant is equivalent\cite{papadakis2015trivial}. Mutation score is the ratio of killed mutants out of all mutants. Mutation coverage aims to identify a minimal test suite that achieves a mutation score of 100\%. This technique has been shown to subsume traditional structural adequacy criteria~\cite{li2009experimental,ammann2017introduction}. Additionally, empirical studies have shown that mutation-based criteria correlate more strongly with fault detection effectiveness than structural criteria such as statement and branch coverage~\cite{chekam2017empirical}.

\subsection{Fault Detection of Traditional Coverage Criteria}

The effectiveness of test adequacy criteria in automated software testing has long been a topic of interest in both academia and industry. As a result, numerous studies have investigated the relationship between adequacy criteria and testing effectiveness across a variety of settings. 

The effectiveness of test adequacy criteria was primarily evaluated on existing developer-written programs under test. At this point, the literature can be divided into two categories: studies that assess adequacy criteria using real faults, and studies that rely on artificially injected faults, such as mutants.

Initially, a number of studies evaluated the effectiveness under real faults. Frankl and Weiss ~\cite{10.1145/120807.120821} compared branch coverage with the all-uses criterion, while Frankl et al~\cite{FRANKL1997235}. later compared all-uses testing with mutation testing. Subsequently, Frankl and Iakounenko~\cite{10.1145/288195.288298} investigated the relationship between all-uses and branch coverage. Although these studies were conducted on relatively small experimental subjects, they consistently reported a positive association between test adequacy and fault detection effectiveness. In particular, the all-uses criterion generally outperformed branch coverage, whereas mutation testing demonstrated the highest fault detection effectiveness among the evaluated criteria. However, these studies did not compare all criteria within a common experimental setting, limiting direct comparisons across techniques. Other studies focused on evaluating a single criterion in isolation~\cite{ciupaetal, Wei2012}. Their findings are consistent with prior work but no comparison between the test adequacy metrics was conducted.

Due to the limited availability of real faults, subsequent research increasingly relied on mutation analysis and artificially injected faults to evaluate adequacy criteria~\cite{6606563, ANAND20131978, 10.1145/2568225.2568278, 10.1145/2568225.2568271}. Most of these studies, however, operated under the \emph{Clean Program Assumption}, whereby faults are injected into an otherwise correct program. Later on, Chekam et al.~\cite{chekam2017empirical} demonstrated that this assumption does not accurately reflect the characteristics of naturally faulty code.

Perhaps the study most closely related to ours is that of Chekam et al.~\cite{chekam2017empirical}. They challenged the \emph{Clean Program Assumption} and evaluated the relationship between test adequacy criteria and fault detection effectiveness using developer-written and automatically generated test suites. Their findings suggest that strong mutation leads to higher fault detection. However, their study predates the era of LLM code generation. Our work revisits this question in the modern era where both code and tests are LLM-generated. 

\subsection{LLM-generated Code and Test}
Recent advances in Large Language Models (LLMs) have changed dramatically how software development tasks are handled. LLMs are used for code completion~\cite{chen2021evaluating, jiang2024surveylargelanguagemodels}, bug finding~\cite{DBLP:journals/pacmpl/LiHZQ24}, program repair~\cite{DBLP:conf/sigsoft/JinSTSLSS23}, and requirements formalization \cite{NEURIPS2022_d0c6bc64}, among many other tasks. 
In particular, LLMs are increasingly being integrated into software agents to automatically generate code to implement new and existing features~\cite{YangJWLYNP24}. 
Some recent works have studied the impact of LLMs non-determinism in the code generation process~\cite{ZhuLLZLJ024}, how issues present in prompts affect the quality of the generated code~\cite{larbi2025prompts}, and explore different kinds of bugs LLM can typically produce~\cite{10.1007/s10664-025-10614-4}.

LLMs are also being used for automated test generation~\cite{konstantinou2026llmbasedtestgenerationtechniques,WangL0J24}. Such approaches rely on traditional test adequacy criteria (branch, statement, and mutation coverage) to evaluate the quality of the generated test suites. Other approaches incorporate these criteria as generation objectives by providing feedback to the LLM. For instance, CoverUp~\cite{abs-2403-16218} aims to maximize statement coverage, while TestSpark~\cite{sapozhnikov2024testspark} and YATE~\cite{konstantinou2025yateroletestrepair} focus on branch coverage. More recently, mutation analysis has also been adopted to guide the generation of fault-revealing tests~\cite{dakhel2024effective,chang2026test}.

 
Although several studies compared human-written code vs LLM-generated code~\cite{reinhart2025llms}, 
 or human-written vs LLM-generated test cases~\cite{vathana2026llm}, to our knowledge, there is no study on the effectiveness and relationship between traditional test adequacy criteria under LLM-generated code and test suites. Our work addresses this gap through a large-scale empirical study.

Vathana et al.~\cite{vathana2026llm} compared the fault detection effectiveness of human-written and LLM-generated test cases. As part of their study, they also examined the relationship between statement and branch coverage and fault detection effectiveness, finding that high coverage does not necessarily imply strong fault detection. Our findings are consistent with these observations. However, their study focused on developer-written code and considered only statement and branch coverage. In contrast, our study focuses on the fault detection effectiveness of test suites generated under different adequacy criteria, including mutation testing, rather than on whether the criteria themselves guarantee high fault detection. 

Liu et al.~\cite{doi.org/10.1002/smr.70034} evaluated the quality of code-generation benchmarks commonly used to assess LLM-based test generation approaches. Their study considered two of the benchmarks used in our work and evaluated the benchmark-provided test suites using statement coverage, branch coverage, and mutation testing. However, their analysis focused on the existing benchmark test suites. In contrast, we investigate the fault detection effectiveness of these adequacy criteria when both the code and the test suites are generated by LLMs. Nevertheless, their findings motivated our decision to augment the benchmark test suites in order to collect additional faulty implementations.

To the best of our knowledge, this is the first large-scale study that systematically evaluates the relationship between test adequacy criteria and fault detection effectiveness on LLM-generated code, including code produced from correct and under-specified prompts.

\section{Research Questions}

Vibe coding introduces a new world for software development. In this world, both code and tests are LLM generated, so we investigate the test adequacy question by analyzing:

\textbf{RQ1 (Fault Detection):} \textit{To what extent do statement, branch, and mutation coverage relate with fault detection and fault triggering in the case of LLM-generated code?}

The answer to this question will provide evidence showcasing the extent to which the coverage criteria are suitable for testing generated code, as done for traditional software. The difference between the fault detection and fault triggering cases provides insight on whether the tests are capable of detecting the faults but fail to actually capture it through a correct assert statement. In such a case, testers will need to manually check and revise the generated test assertions.

Our results found evidence that LLM-generated tests have reasonably strong fault triggering ability but very low fault detection, mainly due to incorrect test oracles. 
Since these (incorrect) test oracles will require manual inspection, introducing an important testing cost factor, we now compare in a cost-effective way the different test criteria: a criterion that requires fewer tests to achieve the same fault detection (or triggering) is preferable over another one that requires more effort. 
Therefore, we ask: 

\textbf{RQ2 (Cost-Efficiency):} \textit{How efficient are statement, branch, and mutation coverage criteria relative to the testing effort they require?}

Recent studies have shown that LLMs  focus on the current implementation rather than the specification when producing test oracles \cite{konstantinou2024llmsgeneratetestoracles,huang2024measuring}. Since in our study we use faulty implementations, it can mislead the LLMs toward producing a significant number of defective test oracles~\cite{konstantinou2024llmsgeneratetestoracles}. We thus investigate whether LLMs can make the LLMs generate correct test oracles, as suggested by previous work \cite{huang2024measuring}, and thus improve their fault detection, when the code implementation is hidden and just the triggering test inputs along with the NL description of the expected behavior are used to generate test oracles. Hence, we ask:

\textbf{RQ3 (Fault Detection and the Oracle Problem):} \textit{To what extent can specification-guided oracle generation improve the fault detection capability of LLM-generated tests?}

\section{Experimental Setup}

This study investigates whether traditional test adequacy criteria provide good guidance in finding faults in LLM-generated code. 
We use state-of-the-art code generation models and benchmark datasets to construct a collection of faults produced by LLMs from natural language descriptions. These faults represent mistakes that may arise during vibe coding, where developers provide natural language prompts and the LLM generates a corresponding implementation. In the following, we first introduce the necessary definitions, then present the selected models and datasets. Next, we describe the process of constructing our benchmark of non-trivial faults, and finally present test and oracle generation. 

\subsection{Definitions}
\label{sec:defs}
\subsubsection{Fault Triggering and Fault Detection} 
These two related concepts capture different stages of how a test exposes a fault.
Triggering measures whether a fault causes any observable behavioral
deviation. Detection measures whether the test's oracle actually
\emph{catches} that deviation. A fault can be triggered yet not detected if
the oracle is incorrect.

\begin{definition}[Fault Trigger]
A fault $f$ is \emph{triggered} by test $t$ if executing $t$ on the faulty
program produces a behavioral outcome that differs from the output of the
correct program $p$:
\[
  \mathrm{triggered}(f,\, t)
  \;\leftrightarrow\;
  \mathrm{output}(f,\, t) \neq \mathrm{output}(p,\, t)
\]
A fault $f$ is triggered by test suite $T$ if at least one test in $T$
triggers it:
\[
  \mathrm{triggered}(f,\, T)
  \;\leftrightarrow\;
  \exists\, t \in T
  : \mathrm{triggered}(f,\, t)
\]
\end{definition}

\begin{definition}[Fault Trigger Rate]
\[
  \mathrm{FTR}(T)
  = \frac{|\{f \in F : \mathrm{triggered}(f,\, T)\}|}{|F|}
\]
\end{definition}

\begin{definition}[Fault Detection]
A fault $f$ is \emph{detected} by test $t$ if $t$ triggers the fault
and the test oracle flags the output as incorrect, i.e., the
oracle captures a discrepancy between the observed output and the expected
behavior:
\[
  \mathrm{detected}(f,\, t)
  \;\leftrightarrow\;
  \mathrm{triggered}(f,\, t)
  \;\wedge\;
  \mathrm{output}(f,\, t) \neq \mathrm{oracle}(t)
\]
A fault $f$ is detected by test suite $T$ if at least one test in $T$
detects it:
\[
  \mathrm{detected}(f,\, T)
  \;\leftrightarrow\;
  \exists\, t \in T
  : \mathrm{detected}(f,\, t)
\]
\end{definition}

\begin{definition}[Fault Detection Rate]
\[
  \mathrm{FDR}(T)
  = \frac{|\{f \in F : \mathrm{detected}(f,\, T)\}|}{|F|}
\]
\end{definition}

Intuitively, when the oracle is perfect, detection
and triggering coincide. But in practice, typically $\mathrm{FDR} \leq \mathrm{FTR}$, since
a weak or missing oracle may allow triggered faults to go undetected.

\subsubsection{Fault Difficulty}

\begin{definition}[Fault Difficulty]\label{fault-difficulty-score}
The difficulty of fault $f$ relative to test suite $T$ is defined as the
complement of the proportion of tests that trigger it:
\[
  \mathrm{difficulty}(f,\, T)
  = 1 -
  \frac{|\{t \in T : \mathrm{triggered}(f,\, \{t\})\}|}{|T|}
\]
where $\mathrm{difficulty}$ ranges from $0$ (trivial: every test triggers the
fault) to $1$ (hard: no test triggers the fault).
\end{definition}

\subsubsection{Mutation Score}

\begin{definition}[Killed Mutant]
A mutant $m$ is \emph{killed} by test suite $T$ if at least one test
$t \in T$ produces output on $m$ that differs from the output on the original
program $p$:
\[
  \mathrm{killed}(m,\, T)
  \;\leftrightarrow\;
  \exists\, t \in T
  : \mathrm{output}(m,\, t) \neq \mathrm{output}(p,\, t)
\]
\end{definition}

\begin{definition}[Mutation Score]
The mutation score is the ratio of killed mutants to all non-equivalent
mutants generated:
\[
  \mathrm{MS}(T)
  =
  \frac{|\{m : \mathrm{killed}(m,\, T)\}|}
       {|\{m : \mathrm{killed}(m,\, T)\}|
        +
        |\{m : \mathrm{surviving}(m,\, T)\}|}
\]
\end{definition}

\subsection{Datasets and Models}

We employ four established python datasets comprising diverse coding tasks. Each benchmark consists of a programming task specified through a prompt that requires the model to generate a code solution. For each task, the benchmark provides a reference implementation along with a set of tests used to assess the correctness of generated solutions. Namely, we used the following benchmarks in our study:

\textit{HumanEval+}\cite{liu2024your}\cite{chen2021evaluating} is a python benchmark to evaluate the LLMs' ability to generate code from a natural language description. It contains 164 programming problems written by humans and an extensive set of test cases. 

\textit{MBPP} \cite{austin2021program} a benchmark of 974 short python programming designed for entry-level difficulty, the test sets were later extended in a followup work\cite{liu2024your}.

\textit{BigCodeBench} \cite{zhuo2025bigcodebench}, given that most code generation benchmarks at the time focused on algorithmic problems, BigCodeBench is a set of 1140 tasks from 139 libraries to evaluate the LLMs' code generation abilities on complex instructions, function calls, and library use. 

\textit{NaturalCodeBench} \cite{zhang2024naturalcodebench} is code benchmark designed to reflect the
complexity and variety of applications in real coding tasks, it contains 402 problems in Python and Java but only 70 of them are open-source. It covers 6 different domains such as algorithms, data science, system administration, and front-end. 

We employ five state-of-the-art large language models to ensure diversity, commercial and open-source from different model families. Specifically, we use: GPT-5-mini and and GPT-4.1-mini from OpenAI, DeepSeek-V4-Flash, Claude Haiku 4.5, and Llama 3.3 Instruct (70B).

\begin{figure}
    \centering
    \includegraphics[width=\columnwidth]{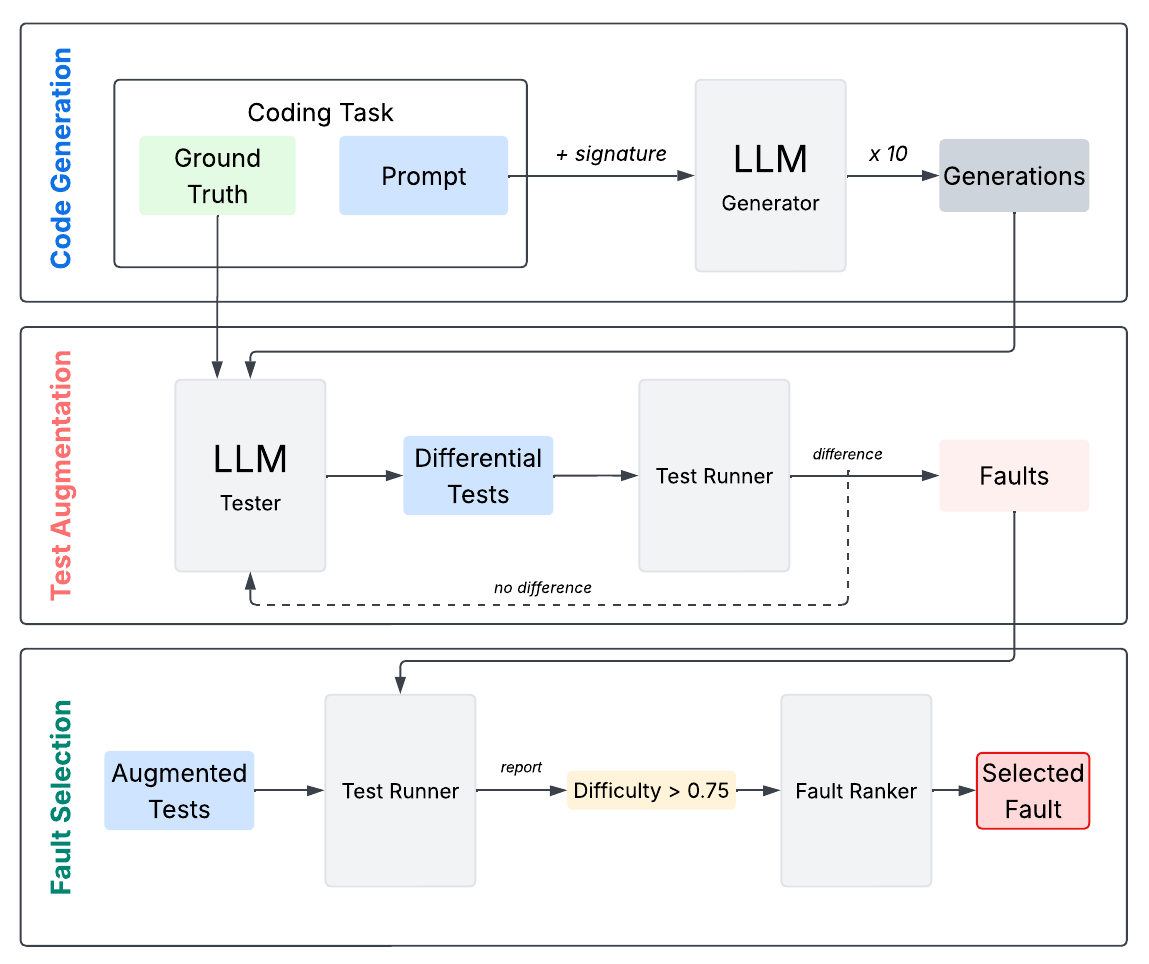}
    \caption{\textbf{Overview of fault collection} Inputs include a task with original and under-specified NL prompts. Prompts are used to produce generations. Test augmentation generates differential test suites to find faulty generations. Faults meeting the threshold are ranked by difficulty and the most difficult one is selected for the task.}
    \label{fig:augmented_tests_quality}
\end{figure}

\subsection{Non-Trivial Faults Collection}


\subsubsection{Code Generation}


From each of the datasets, we take the natural language task descriptions that will be used for prompting the models. 
In addition, we consider the prompt variations proposed for these datasets by Larbi et al.\cite{larbi2025prompts,akli2026defective}, where different kinds of defects (ambiguous, incomplete, under-specified, etc) were injected, to increase the likelihood and diversity of the faulty implementations generated. 

Hence, we instruct the LLMs to generate an implementation for both the original and defective prompts. We repeat this process 10 times for each LLM, using a high temperature (0.8) to diversify the generation process.
In total, we generated 233,300 implementations, determined by the benchmark size × 10 generations × 2 prompt variants (original and underspecified) × 5 models, as summarized in Table~\ref{tab:generations}.

\begin{table}[h]
\centering
\caption{Total number of generated implementations for each benchmark}
\label{tab:generations}
\resizebox{\columnwidth}{!}{
\begin{tabular}{lcccc}
\toprule
Model & HumanEval & BCB & MBPP & NCB \\
\midrule
N° generations & 16,400 & 113,000 & 97,400 & 6,500 \\
\bottomrule
\end{tabular}
}
\end{table}

In our study, the reference solution provided by the dataset serves as our ground truth. 
Our goal is to collect faulty implementations, and thus, we proceed to run the generated implementations and the reference solution on the same given test suite. 
If the generated implementation follows a different execution path (i.e., exhibits faulty behavior), we keep it. 
Since recent works~\cite{doi.org/10.1002/smr.70034} have shown limitations in the provided benchmarks' test suites, not rigorous enough to expose faulty implementations, we proceed to augment the corresponding test suites. 

\subsubsection{Test Augmentation}


For each generated implementation, we use an LLM to generate additional test cases that are designed to identify behavioral differences between the generated implementation and the ground truth. The objective of this process is to obtain tests that can expose even subtle incorrect behavior.
A differential test may fail on both the reference implementation and the generated implementation; however, this does not necessarily indicate that the generated implementation is faulty. To determine whether the generated implementation indeed contains a fault, we compare the outputs produced by the two implementations for the corresponding test input. If the outputs differ, we conclude that the generated implementation exhibits behavior different from the reference solution and classify it as faulty. Otherwise, the generated implementation is discarded. 
Hence, we identify 73,785 valid faulty implementations of the 233,300 generated.  

\begin{figure}[h]
    \centering
    \includegraphics[width=\columnwidth]{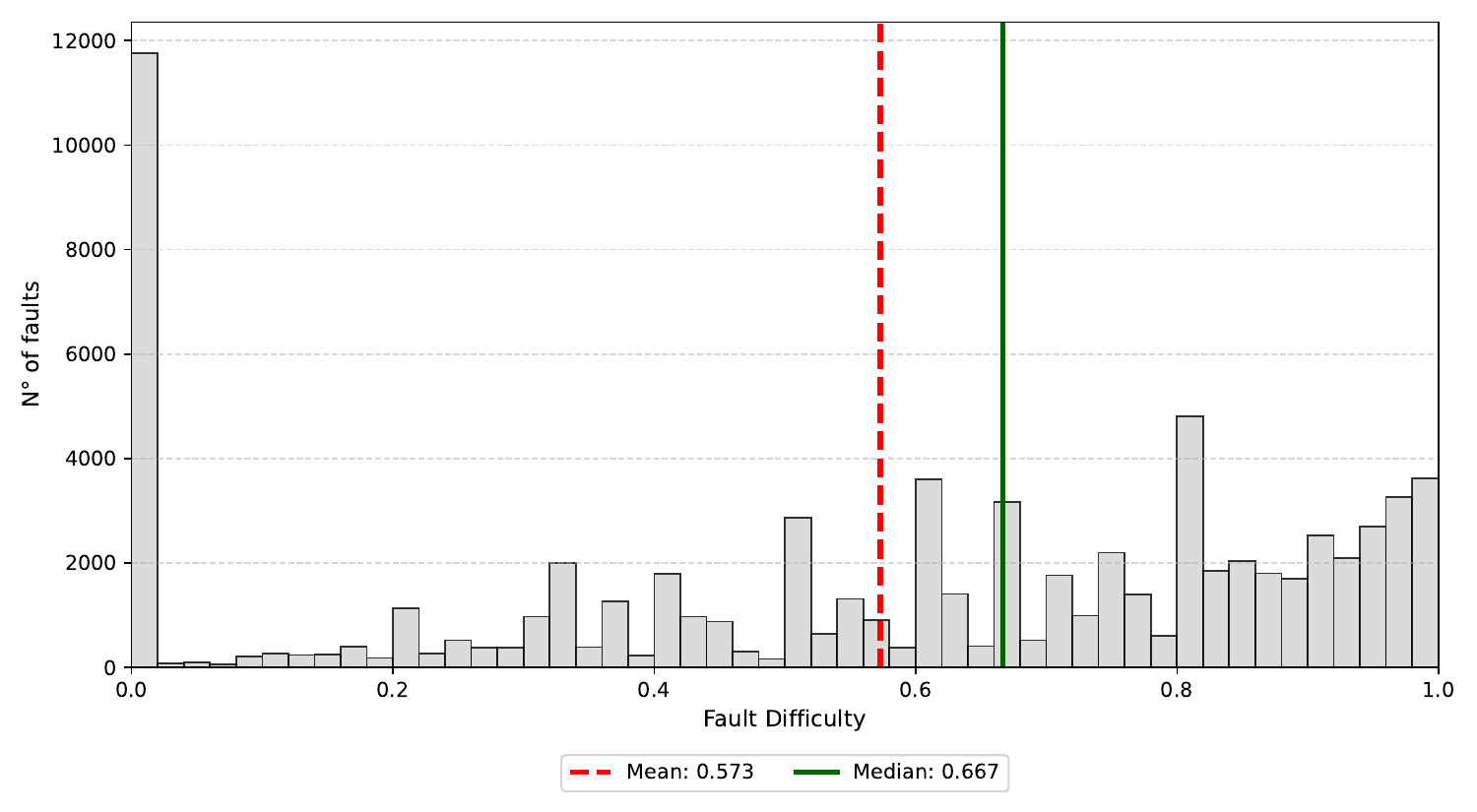}
    \caption{Distribution of fault difficulty across all faults. The histogram illustrates the proportion of faults at each difficulty level, ranging from easy (difficulty = 0) to difficult (difficulty $\sim$ 1)}
    \label{fig:difficulty}
    \vspace{-1em}
\end{figure}

Figure \ref{fig:difficulty} shows the distribution of fault difficulty across 73,785 faulty generations. The distribution reveals that approximately 16\% of the faults are extremely simple to find, being killed by more than 99\% of the tests. The median difficulty is 0.66, which is still considered easy. Since a task may have multiple non-trivial faulty versions (on average 2.92), we pick the most difficult fault per task.

\subsubsection{Fault Selection}

The previous steps yield a collection of faulty implementations. However, not all faults are equally valuable for evaluation. Some faults are trivial i.e. they are exposed by a large number of test cases, while others are more subtle and require stronger test suites to be detected. In the context of testing, there is no much of value in evaluating test criteria against trivial faults as any ad-hoc testing process would reveal them. To this end, we opted for selecting the most challenging ones, similar to the practice adopted by previous studies \cite{ChekamPBTS20, DoER05}. To focus the evaluation on the most challenging faults, we rank faulty implementations according to their fault difficulty score (described in definition~\ref{fault-difficulty-score}) achieved on the augmented tests suites. 

After assigning a difficulty score to each fault, we filter the faulty implementations as follows:

\begin{enumerate}
    \item We discard implementations with a fault difficulty score lower than 0.75 (the same threshold has been taken by other studies \cite{ChekamPBTS20, DoER05}). This threshold ensures that the final dataset contains only faults that are detected by fewer than 25\% of the available test cases.

    \item For tasks with multiple faulty implementations, we retain only the implementation with the highest fault difficulty score, corresponding to the most challenging-to-detect fault. Thus, each task contributes at most one faulty implementation to the final dataset.

\end{enumerate}
\begin{table}[t]
\centering
\caption{Number of non-trivial faults collected per benchmark and model}
\label{tab:faulty-implementations}
\begin{tabular}{lccccc}
\toprule
Model & HE & BCB & MBPP & NCB & Total \\
\midrule
GPT-5-Mini   &  84 &  845 & 355 & 16 &   1300  \\
GPT-4.1-Mini  & 91 &  608 & 280 & 14 &  993  \\
Claude-Haiku 4.5 & 105 & 744 & 294 & 20 &  1163 \\
Deepseek-v4-flash  & 115  & 912 & 372 & 22 &   1421 \\
Llama 3.3-70B-Instruct & 125 & 816 & 229 & 19  &  1189 \\
\midrule
Total   & 520 & 3925 & 1530 & 91 &  \textbf{6066} \\
\bottomrule
\end{tabular}
\end{table}

Table~\ref{tab:faulty-implementations} reports the number of non-trivial faults collected after applying the previous steps, broken down by benchmark and model. Across the four benchmarks and five models, we obtain a total of \emph{6,066} faulty implementations for our study.


\subsection{LLM-based Test Generation}
\begin{figure}[t]
    \centering
    \includegraphics[width=\columnwidth]{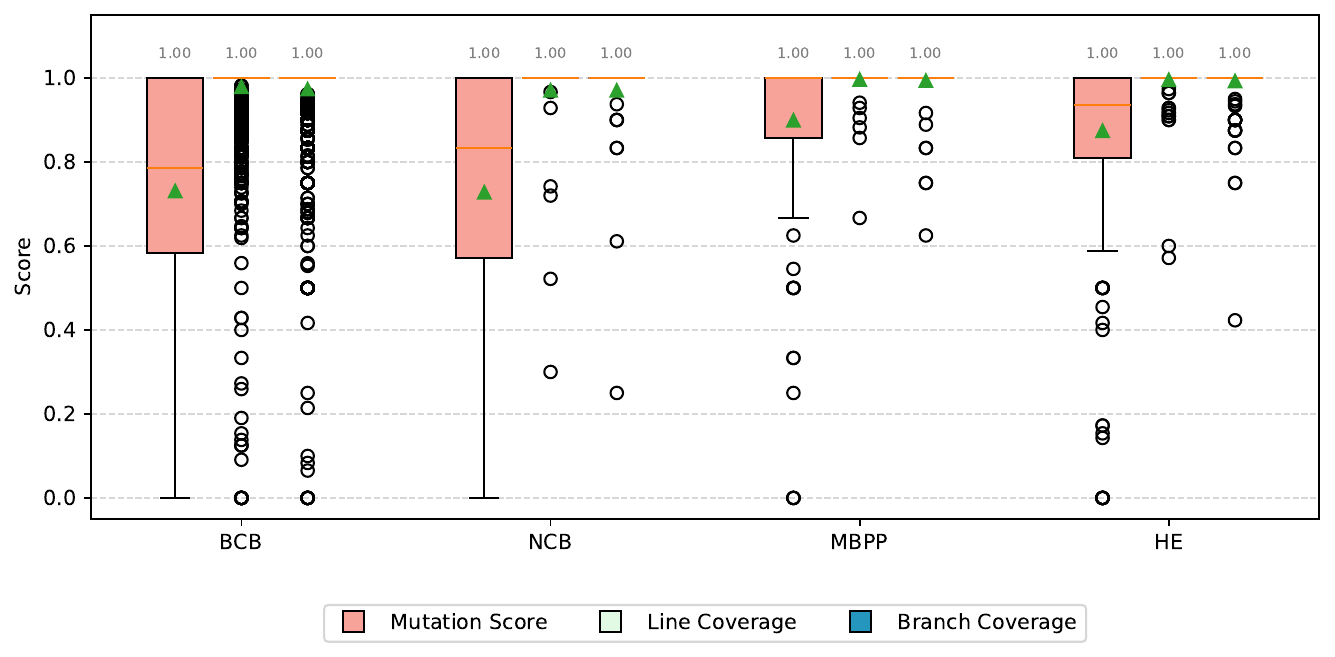}
    \caption{Distribution of the Coverage Scores of the LLM-Generated Tests} 
    \label{fig:test_suites_coverage}
    \vspace{-1em}
\end{figure}

To conclude with our experimental setup, we generate a large pool of test cases from which we will later sample test suites w.r.t. the test adequacy criteria under analysis. 
Particularly, we use \emph{LLM-Plain}~\cite{konstantinou2026llmbasedtestgenerationtechniques}, a simple LLM-based test generation approach that just relies on the capabilities of LLMs (i.e. no coverage information is considered during the test pool generation). 
Despite its simplicity, LLM-Plain has been shown to outperform several recent state-of-the-art test generation techniques. 
We run LLM-Plain using its default configuration, and generated a total of 4872 tests for HumanEval, 30450 for BCB, 24007 for MBPP, and 1256 for NCB. 
Figure~\ref{fig:test_suites_coverage} summarizes the mutation score, statement and branch coverage, achieved by the LLM-generated test suites.

\begin{table}[h]
\centering
\caption{FDR and FTR of the LLM-generated-tests per benchmark : models aggregated}
\label{tab:FTR_FDR}
\resizebox{\columnwidth}{!}{
\begin{tabular}{lcccc}
\toprule
Model & HumanEval & BCB & MBPP & NCB \\
\midrule
Fault Trigger & 0.316 & 0.623 & 0.376 & 0.532 \\
Fault Detection & 0.032 & 0.043 & 0.027 & 0.080 \\
\bottomrule
\end{tabular}
}
\end{table}

\subsection{NL Specification-guided Oracle Generation}\label{oracle-gen}
Recent evidence~\cite{konstantinou2024llmsgeneratetestoracles} that, when exposed to faulty code, the LLM-oracle generation process is biased toward the observed behavior of the faulty program, rather than the intended behavior of the task. 
This can potentially bias our evaluation and hinder the fault detection capabilities of the test adequacy criteria. 


To address this, we use the natural language specifications provided in the benchmarks and run a specification-guided oracle generation. First, we identify tests that successfully trigger discrepancies between the faulty and ground-truth implementations. From each such test, we extract the test prefix containing the test inputs, and prompt the LLM using the NL description to complete the test prefix by generating an oracle that asserts the behavior expressed in the description. 
In this process, the current faulty implementation is not included in the prompt. 
Finally, in cases where the resulting test cases fail due to runtime errors, we iteratively re-prompt the LLM with the execution feedback. This repair process is repeated up to five times, aiming to produce syntactically correct tests.

\section{Experimental Protocol}
We simulate a testing scenario where a tester selects tests to maximize a given coverage criterion (statement, branch, mutation), and then measure the obtained test suites' fault triggering rate and fault detection rate (see Section~\ref{sec:defs}). 

Given a faulty implementation $f$ and the pool of generated tests $TS_f$, we iteratively select a random test $t \in TS_f$ and check whether it increases the coverage according to the target criterion $C$. If it does, we keep it; otherwise, we discard it and select another test. 
The process stops when the coverage of the sampled test suite $TS_{f}^{i} \subseteq TS_f$ achieve same coverage as the test pool $TS_f$ (i.e. $C(TS_f) = C(TS_{f}^{i})$). 
This simulation is typical in the literature for evaluating fault detection capabilities\cite{titcheu2020selecting, ma2020commit}.

Since the process is randomized, we repeat it 100 times per fault per criterion. In total, for each coverage criteria, we sampled 52,000 test suites for HumanEval, 392,500 for BCB, 153,000 for MBPP, and 9100 for NCB.

To address RQ1, for each sampled test suite, we study whether the faults are triggered and detected. Thus, we compute the fault trigger rate (FTR) and fault detection rate (FDR) as defined in Section~\ref{sec:defs}, across the 100 sampled test suites for each criterion, and report the mean values for every benchmark–model combination. Additionally, we conduct a statistical analysis (Wilcoxon test and Vargha-Delaney effect size $\hat{A}_{12}$) to assess whether the differences observed between criteria are statistically significant.

The goal of RQ2 is to perform a cost-effective comparison between the different test adequacy criteria. 
Hence, by reusing the same simulation as before, we measure the effort of a test suite as the number of tests executed, and compare their effectiveness measured as the fault triggering rate (FTR) and fault detection rate (FDR). 
After a normalization across faults and runs, we plot the progression of FTR and FDR as the effort increases from 0 to 100\%. 

Recent evidence has shown that faulty code can mislead LLMs when producing test oracles, thus hindering their fault detection capabilities~\cite{konstantinou2024llmsgeneratetestoracles,huang2024measuring}. 
RQ3 aims to explore whether providing the LLM just with distinguishing inputs and the expected behavior (NL prompt), ignoring the current (faulty) implementation, can improve  the test oracle generation, reducing the need for manual oracle validation, and ultimately increasing the fault detection rate.

To answer this question, we start from the LLM-generated test suites, which we refer to as \emph{base} suites. We collect the triggering tests from these suites and apply the oracle generation procedure described in Section~\ref{oracle-gen}. We then execute the resulting test suites, referred to as \emph{corrected} suites, and record the number of triggered and detected faults. We compare the corrected suites against the base suites to measure the change in both fault triggering and fault detection.

\section{Results}

\subsection{RQ1 : To what extent do statement, branch, and mutation coverage differ in fault detection and fault triggering for LLM-generated code?}

Table~\ref{tab:rq1-results} summarizes the mean fault triggering rate (FTR) and fault detection rate (FDR) achieved by statement, branch, and mutation coverage across all benchmark-model combinations. The results indicate that the choice of test adequacy criterion has only a limited influence on the ability of the generated test suites to trigger and detect faults. Across the evaluated benchmarks, no criterion consistently outperforms the others: mutation generally achieves the best results on BigCodeBench, branch coverage performs best on MBPP, and statement coverage on NaturalCodeBench. Although the superior criterion changes across benchmarks and models, the differences remain small, confirming that the impact of the criterion is limited. 

The statistical analysis in table \ref{tab:rq1-size-effect} reinforce this finding, while some comparison are statistically significant, the size effect remains close to 0.5 which means the differences are small. This confirms that the practical differences between mutation score, branch coverage, and statement coverage are limited.

Although the test suites studied were carefully sampled to maximize statement, branch, and mutation coverage, they exhibit moderate fault triggering and low fault detection. This indicates that  satisfying these traditional test adequacy criteria does not guarantee exposing the LLM generated faults, even the ones collected that are relatively simple (failing up to 25\% of the augmented tests suites).

Beyond the differences between the criteria, we observe an important limitation of LLM-generated tests. While across benchmarks and models, LLM-generated tests can trigger between 25\% and 78\% of the faults (FTR), they can rarely detect thes (FDR between 0\% and 10\%).  
This indicates that LLMs are capable of generating inputs that distinguish faulty from correct behavior but are unable to assert the correct behavior.

\begin{tcolorbox}[colback=gray!5, colframe=gray!40, rounded corners, boxsep=8pt]
  \textbf{Finding 1:} High coverage levels (mutation score, statemnt, and branch coverage) do not guarantee fault trigger or fault detection. This indicates that in LLM-generated code, traditional test adequacy criteria are not reliable indicators of fault detection.
\end{tcolorbox}

 \begin{tcolorbox}[colback=gray!5, colframe=gray!40, rounded corners, boxsep=8pt]
  \textbf{Finding 2:} LLM-generated test suites are more successful at triggering the faults than detecting them. While the generated tests exercise the faulty path, they fail to assert the correct behavior. 
\end{tcolorbox}

\begin{table*}[t]
\centering
    \caption{Fault Trigger Rate (FTR) and Fault Detection Rate (FDR) across five LLM fault models and four benchmarks comparing the three coverage criteria (mutation score, branch 
coverage, statement coverage). Underlined values indicate the 
best-performing criterion for each metric. Results are averaged over 100 iterations.}
    \input{rq1_summary}
    \label{tab:rq1-results}
\end{table*}

\begin{table*}[t]
    \vspace{0.5em}
    \centering
    \caption{Pairwise comparison of the criteria using the Wilcoxon test(p-value) and Vargha–Delaney $\hat{A}_{12}$ effect size across five models and four benchmarks. (\checkmark) denotes rejection of the null hypothesis (p\_val $<$ 0.05), while ($\times$) denotes failure to reject the null hypothesis (p\_val $\geq$ 0.05). $\hat{A}_{12}$ values indicate the magnitude and direction of the observed effect.}
\vspace{0.5em}
    \input{rq1_stats}
    
    \label{tab:rq1-size-effect}
\end{table*}

\subsection{RQ2 : How efficient are statement, branch, and mutation coverage criteria relative to the testing effort they require?}

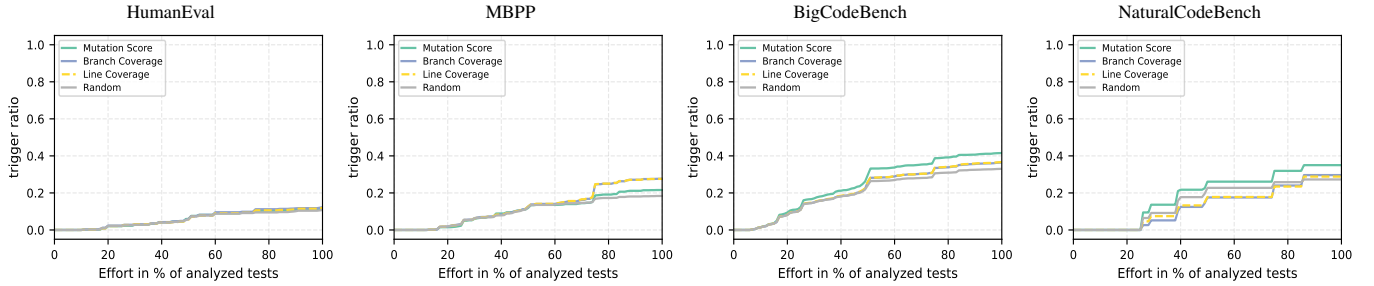
\begin{figure*}[h]
    \vspace{0.6em}
    \centering
    \input{rq2_grid}
        \caption{Cost-efficiency: Fault triggering rate (FTR) evolution with respect to the number of tests analyzed. The curves represent the mean FTR over 100 simulations for mutation, branch coverage, statement coverage, and the random baseline. Each plot represents the results obtained using GPT-4.1-mini for one of the benchmark : HumanEval, MPBB, BigCodeBench, and NaturalCodeBench.}
    \label{fig:rq2-cost-effectiveness}
    \vspace{0.6em}
\end{figure*}

Figure~\ref{fig:rq2-cost-effectiveness} presents the fault triggering rate (FTR) results obtained for each benchmark showing one representative LLM GPT-4.1-mini per benchmark. 
We observe the same trends using other LLMs but due to the lack of space, we only show one model example and provide detailed results in our replication package. Similarly, we do not include the FDR results in the figure because the overall detection rates are low and the observed differences are barely noticeable.

As can be observed from the line plots, the different criteria exhibit similar behavior at the lowest effort levels across all benchmarks. For all benchmarks except HumanEval, the differences become more visible as we increase the effort by selecting more tests to analyze. On BigCodeBench and NaturalCodeBench, mutation consistently achieves the highest FTR under the same budget, while on MBPP, we observe that branch and statement coverage start to outperfom mutation score starting from about 70\% effort. In practice, this means that considerable effort is required before the developer/tester can see the impact of the testing objective. 

Additionally, we observe slight and sometimes no differences between branch and statement coverage, likely because most tests that succeed in covering lines also cover branches. In NaturalCodeBench, we even notice that as the effort increases, branch and statement coverage do not show differences from the random baseline either, similarly, we observe no difference for any of the criteria to the random baseline. This is likely because the selected tests already achieve high coverage, indicating that in such cases they provide no additional benefits. 

 \begin{tcolorbox}[colback=gray!5, colframe=gray!40, rounded corners, boxsep=8pt]
  \textbf{Finding 3:} When considering the human-effort required to analyse the tests, traditional test adequacy criteria provide a weak prioritization signal with little improvement against random sampling. 
\end{tcolorbox}

\subsection{RQ3 : To what extent can specification-guided oracle generation improve the fault detection capability of LLM-generated tests?}

Table \ref{tab:rq3-specs-effect} presents the fault detection rates of the generated test suites before and after specification-guided oracle generation. For each LLM, benchmark, and test adequacy criterion (mutation, line coverage, and branch coverage), the table reports the fault detection rate of the original generated test suite (FDR) and the fault detection rate after replacing the generated assertions with specification-guided oracles (FDR+).

As expected, there is no observable difference between FDR and FDR+ for branch and statement coverage, this is because oracle generation modifies only the assertions while leaving the test prefix the same. In contrast, mutation consistently improves across almost all benchmarks and models after replacing the original assertions with the generated oracles. The improvement can reach up to 27\%, with the best case jumping from 1\% to 28.6\%. 
This indicates that, on the one hand, the generated test prefixes already exercise relevant portion of the code, and on the other hand, faulty code can lead to incorrect test oracle generation, suggesting that better oracles can be obtained when code is hidden. 

Despite this improvement, the overall fault detection remains low and even when provided with fault-triggering inputs and explicit behavioral specifications, the corrected test suites detect at most approximately 30\% of faults. These findings indicate that LLMs are still unable to generate sufficient and effective test cases. Consequently, simply generating test suites with LLMs is not sufficient to reliably catch faults. Human involvement remains essential throughout the testing process to analyze the generated tests and validate the oracle, while explicitly providing the natural language description of the intended behavior might slightly help, considerable effort is still required to ensure the quality of the tests.  
 \begin{tcolorbox}[colback=gray!5, colframe=gray!40, rounded corners, boxsep=8pt]
  \textbf{Finding 4:} NL Specification-guided oracle generation improves fault detection but remains insufficient to produce reliable fault-revealing tests. Test analysis effort is still required to validate the oracles. 
\end{tcolorbox}

\begin{table*}[t]
    \centering
    \caption{Fault Detection Rate comparison between the base test suite (FDR) and the enhanced test suite (FDR+) after generating test oracles from natural language specifications. Results are reported across five LLMs, four benchmarks, and three coverage criteria (Mutation Score, Branch Coverage, and Statement Coverage). Arrows indicate the change in FDR from FDR to FDR+ (↑ improvement, ↓ degradation, no arrow = no change).}

    \input{rq3_FD}
    \label{tab:rq3-specs-effect}
\end{table*}

\section{Threads to Validity}



Construct Validity: 
LLMs are inherently non-deterministic, where repeated queries may yield different outputs. To mitigate this, we repeated our code generation task 10 times, with 5 different LLMs, generated a large number of implementations, and run the simulation 100 times, ensuring results reflect a distribution of behaviors rather than a single observation. 

Another threat relates the LLMs-generated faults may not fully represent real-world defects. We mitigated this by filtering trivial faults, retaining only the most challenging ones to focus on faults that can plausibly escape standard testing.

Internal Validity: Different LLMs can interpret differently the original and under-specified NL prompts. We try to mitigate this by using established benchmarks (HumanEval+, MBPP, BigCodeBench, NaturalCodeBench) with standardised task descriptions. Our protocol also aimed at collecting diverse faults by employing defective task descriptions developed by previous work. 

Additionally, our fault detection relies on comparing generated program behavior against a reference implementation. Any defect in the reference solution could lead to misclassifications of faulty versus correct implementations. We mitigated this by augmenting the test suites to gain confidence in the distinction between correct and faulty behaviors. 

External Validity:  
Our study focuses exclusively on python functions and thus results may not generalize to other languages or more complex systems. 
Although we use five LLMs across four benchmarks, covering a representative range of modern models and tasks, results may not extend to other models and datasets.

\section{Conclusion}
In this paper, we investigated the effectiveness of traditional test adequacy criteria for assessing the fault-detection capability of test suites generated by large language models in a fully LLM-driven software development setting. Our empirical study encompassed 6,066 faults, five mainstream LLMs, and four Python coding benchmarks. The results consistently demonstrate that traditional adequacy metrics—namely statement coverage, branch coverage, and mutation testing—are weak predictors of the ability of LLM-generated test suites to identify faults in LLM-generated code. While these criteria have long been regarded as reliable indicators of test effectiveness in traditional software engineering, our findings suggest that they provide limited guidance in emerging LLM-driven development workflows.

A particularly noteworthy finding of the study is that LLM-generated tests often exercise faulty behavior and expose behavioral differences between faulty and correct implementations, yet still fail to detect these faults due to weak, incomplete, or incorrect test oracles. This observation indicates that the primary challenge is no longer exercising relevant behaviors, but rather reliably specifying and verifying the expected outcomes. Consequently, despite recent progress in LLM-based test generation, human involvement remains essential for validating the correctness of test oracles.

To the best of our knowledge, this is the first empirical study to examine the relationship between traditional test adequacy criteria and fault-detection effectiveness in a fully LLM-driven software development context. While our results do not provide definitive answers regarding how test effectiveness should be measured in this new paradigm, they offer strong empirical evidence that coverage- and mutation-based adequacy criteria are poor indicators of fault-detection capability for LLM-generated tests. These findings highlight the need for further research to better understand how test effectiveness should be assessed in LLM-driven development and whether new adequacy criteria are required to support this emerging software engineering paradigm.

Our study opens several directions for future research. First, there is a need to develop adequacy criteria that effectively target and predict the fault-detection capabilities of tests generated for LLM-produced code. Second, future criteria should place greater emphasis on the quality of test oracles, given that the primary challenge lies in correctly identifying and asserting faulty behavior rather than exercising it. Third, automated techniques are needed to detect and mitigate underspecification in task descriptions and prompts ultimately limiting fault-detection effectiveness. More broadly, future research should explore how testing methodologies and evaluation criteria can be redesigned to better align with the unique characteristics of fully LLM-driven software development workflows.

\bibliographystyle{IEEEtran}
\bibliography{citations.bib}

\end{document}

%% file: rq1_summary.tex
\footnotesize
\begin{tabular*}{\textwidth}{@{\extracolsep{\fill}}llcccccc@{}}
\toprule
& & \multicolumn{2}{c}{Mutation} & \multicolumn{2}{c}{Branch} & \multicolumn{2}{c}{Statement} \\
\cmidrule(lr){3-4} \cmidrule(lr){5-6} \cmidrule(lr){7-8}
Benchmark & Model & FTR & FDR & FTR & FDR & FTR & FDR \\
\midrule
\multirow{5}{*}{HumanEval}
  & gpt-5-mini & 0.393 & 0.000 & \underline{0.450} & 0.000 & 0.385 & 0.000 \\
  & gpt-4.1-mini & 0.111 & 0.020 & \underline{0.161} & \underline{0.026} & 0.120 & 0.019\\
  & claude-haiku-4-5 & \underline{0.137} & \underline{0.021} & 0.103 & 0.010 & 0.088 & 0.010\\
  & Llama-3.3-70B-Instruct & \underline{0.138} & \underline{0.028} & 0.123 & 0.011 & 0.105 & 0.010\\
  & deepseek-v4-flash & 0.056 & 0.000 & \underline{0.082} & \underline{0.006} & 0.070 & 0.003 \\
\midrule
\multirow{5}{*}{MBPP}
  & gpt-5-mini & 0.595 & 0.002 & \underline{0.796} & \underline{0.004} & 0.695 & 0.004 \\
  & gpt-4.1-mini & 0.175 & 0.005 & \underline{0.432} & \underline{0.009} & 0.217 & 0.004\\
  & claude-haiku-4-5 & 0.225 & \underline{0.011} & \underline{0.440} & 0.011 & 0.281 & 0.007\\
  & Llama-3.3-70B-Instruct & 0.149 & 0.012 & \underline{0.403} & \underline{0.022} & 0.238 & 0.017\\
  & deepseek-v4-flash & 0.214 & \underline{0.009} & \underline{0.428} & 0.005 & 0.272 & 0.004\\
\midrule
\multirow{5}{*}{BigCodeBench}
  & gpt-5-mini & 0.774 & \underline{0.010} & \underline{0.814} & 0.007 & 0.793 & 0.008 \\
  & gpt-4.1-mini & 0.436 & \underline{0.019} & \underline{0.437} & 0.014 & 0.377 & 0.012\\
  & claude-haiku-4-5 & \underline{0.381} & \underline{0.024} & 0.286 & 0.007 & 0.245 & 0.005 \\
  & Llama-3.3-70B-Instruct & \underline{0.269} & \underline{0.030} & 0.226 & 0.018 & 0.210 & 0.018 \\
  & deepseek-v4-flash & \underline{0.473} & \underline{0.013} & 0.434 & 0.010 & 0.386 & 0.010\\
\midrule
\multirow{5}{*}{NaturalCodeBench}
  & gpt-5-mini & 0.545 & 0.000 & \underline{0.625} & 0.000 & \underline{0.625} & 0.000 \\
  & gpt-4.1-mini & 0.487 & 0.027 & 0.452 & \underline{0.044} & \underline{0.545} & 0.04 \\
  & claude-haiku-4-5 & 0.336 & 0.001 & 0.234 & \underline{0.008} & \underline{0.343} & 0.007\\
  & Llama-3.3-70B-Instruct & 0.254 & 0.080 & \underline{0.483} & \underline{0.110} & 0.255 & 0.055\\
  & deepseek-v4-flash & 0.308 & 0.046 & \underline{0.382} & \underline{0.082} & 0.338 & 0.055 \\
\bottomrule
\end{tabular*}

%% file: rq1_stats.tex
\footnotesize
\begin{tabular*}{\textwidth}{@{\extracolsep{\fill}}llccc@{}}
\toprule
& & Mutation vs. Statement & Mutation vs. Branch & Statement vs. Branch\\
\midrule
\multirow{5}{*}{HumanEval}
  & gpt-5-mini & p\_val=0.54 $\times$ & p\_val=0.53 $\times$ & p\_val= 0.5 $\times$ \\
  & gpt-4.1-mini & p\_val=0.49 $\times$ & p\_val=0.45 $\times$ &  p\_val=0.50 $\times$ \\
  & claude-haiku-4-5 & p\_val=0.02 $\hat{A}_{12}$=0.54 \checkmark & p\_val=0.1 $\times$ & p\_val=0.26 $\times$ \\
  & Llama-3.3-70B-Instruct & p\_val=0.48 $\times$ & p\_val=0.51 $\times$ & p\_val=0.48 $\times$  \\
  & deepseek-v4-flash & p\_val=0.54 $\times$ & p\_val=0.50 $\times$ & p\_val=0.47 $\times$ \\
\midrule

\multirow{5}{*}{MBPP}
  & gpt-5-mini &  p\_val=0.00 $\hat{A}_{12}$=0.45 \checkmark & p\_val=4e-6 $\hat{A}_{12}$=0.44 \checkmark & p\_val=0.01 $\hat{A}_{12}$=0.48 \checkmark \\
  
  & gpt-4.1-mini & p\_val=0.01 $\hat{A}_{12}$=0.46 \checkmark & p\_val=1e-4 $\hat{A}_{12}$=0.42 \checkmark  &  p\_val=0.12 $\times$ \\
  
  & claude-haiku-4-5 & p\_val=2e-3 $\hat{A}_{12}$=0.47 \checkmark   & p\_val=5e-4 $\hat{A}_{12}$=0.45 \checkmark & p\_val=1e-2 $\hat{A}_{12}$=0.46 \checkmark \\
  
  & Llama-3.3-70B-Instruct & p\_val=7e-6 $\hat{A}_{12}$=0.41 \checkmark & p\_val=6e-6 $\hat{A}_{12}$=0.35 \checkmark & 
  p\_val=0.66 $\times$ 
  \\
  
  & deepseek-v4-flash & p\_val=1e-4 $\hat{A}_{12}$=0.46 \checkmark & p\_val=0.0 $\hat{A}_{12}$=0.40 \checkmark & 
  p\_val=0.36 $\times$ 
  \\
\midrule
\multirow{5}{*}{BigCodeBench}
  & gpt-5-mini & p\_val=0.11 $\times$ & p\_val=0.03 $\hat{A}_{12}$=0.49 \checkmark &  p\_val=7e-5 $\hat{A}_{12}$=0.48 \checkmark\\

  & gpt-4.1-mini & p\_val=5e-6 $\hat{A}_{12}$=0.55 \checkmark & p\_val=2e-4 $\hat{A}_{12}$=0.55 \checkmark & p\_val=0.18 $\times$
  \\
  & claude-haiku-4-5 & p\_val=0.00 $\hat{A}_{12}$=0.58 \checkmark & 
  p\_val=0.00 $\hat{A}_{12}$=0.58 \checkmark & 
  p\_val=2e-4 $\hat{A}_{12}$=0.45 \checkmark
  \\
  & Llama-3.3-70B-Instruct &
    p\_val=1e-6 $\hat{A}_{12}$=0.53 \checkmark & 
      p\_val=0.00 $\hat{A}_{12}$=0.54 \checkmark &
    p\_val=1e-4 $\hat{A}_{12}$=0.47 \checkmark \\
  & deepseek-v4-flash & 
    p\_val=0.00 $\hat{A}_{12}$=0.56 \checkmark &
    p\_val=0.00 $\hat{A}_{12}$=0.57 \checkmark &
    p\_val=1e-4 $\hat{A}_{12}$=0.47 \checkmark \\
\midrule
\multirow{5}{*}{NaturalCodeBench}
  & gpt-5-mini & p\_val=0.18 $\times$ & p\_val=0.18 $\times$ & p\_val=1.00 $\times$ \\
  & gpt-4.1-mini & p\_val=0.5 $\times$ & p\_val=0.45 $\times$ &  p\_val=0.5 $\times$ \\
  & claude-haiku-4-5 & p\_val=0.5 $\times$ & p\_val=0.5 $\times$ & p\_val=0.45 $\times$ \\
  & Llama-3.3-70B-Instruct & p\_val=0.58 $\times$ & p\_val=0.50 $\times$ & p\_val=0.61 $\times$ \\
  & deepseek-v4-flash & p\_val=0.52 $\times$ & p\_val=0.50 $\times$ & p\_val=0.46 $\times$ \\
\bottomrule
\end{tabular*}

%% file: rq2_grid.tex
\newlength{\rqpanelwidth}
\newlength{\rqpanelheight}
\setlength{\rqpanelwidth}{0.24\textwidth}
\setlength{\rqpanelheight}{1.3in}
\newcommand{\rqplaceholder}{%
  \fbox{\rule{0pt}{\rqpanelheight}\rule{\rqpanelwidth}{0pt}}%
}
\newcommand{\rqcolheader}[1]{%
  \multicolumn{1}{c}{\scriptsize #1}%
}
{\setlength{\tabcolsep}{2pt}
\renewcommand{\arraystretch}{1}
\begin{tabular}{c c c c}
\rqcolheader{HumanEval} &
\rqcolheader{MBPP} & 
\rqcolheader{BigCodeBench} &
\rqcolheader{NaturalCodeBench} \\[2pt]
\includegraphics[width=\rqpanelwidth,height=\rqpanelheight]{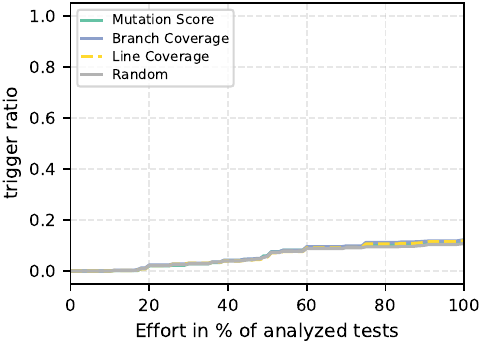} &
\includegraphics[width=\rqpanelwidth,height=\rqpanelheight]{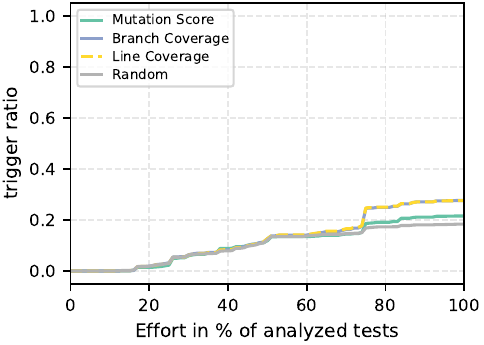}  &
\includegraphics[width=\rqpanelwidth,height=\rqpanelheight]{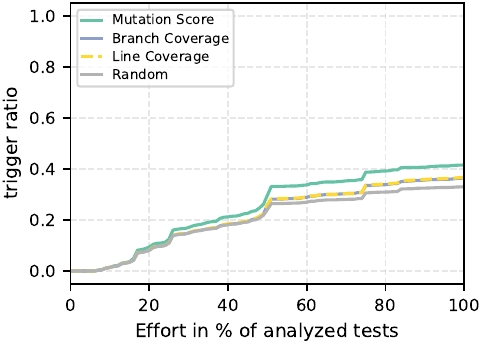} &
\includegraphics[width=\rqpanelwidth,height=\rqpanelheight]{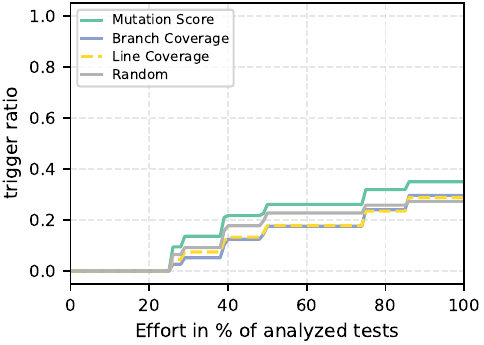} \\[4pt]
\end{tabular}}

%% file: rq3_FD.tex
\footnotesize
\begin{tabular*}{\textwidth}{@{\extracolsep{\fill}}llcccccc@{}}
\toprule
& & \multicolumn{2}{c}{Mutation} & \multicolumn{2}{c}{Branch} & \multicolumn{2}{c}{Statement} \\
\cmidrule(lr){3-4} \cmidrule(lr){5-6} \cmidrule(lr){7-8}
Benchmark & Model & FDR & FDR+ & FDR & FDR+ & FDR & FDR+ \\
\midrule
\multirow{5}{*}{HumanEval}
  & gpt-5-mini & 0.0 & 0.095 \deltaup{} & 0.0 & 0.0 & 0.0 &  0.0\\
  
  & gpt-4.1-mini & 0.020 & 0.044 \deltaup{} & 0.026 & 0.026& 0.019 & 0.019\\
  
  & claude-haiku-4-5 & 0.021 & 0.028 \deltaup{} & 0.010 & 0.010 & 0.010 & 0.010 \\
  
  & Llama-3.3-70B-Instruct & 0.028 & 0.056 \deltaup{} & 0.011 & 0.011 & 0.010 & 0.010\\
  
  & deepseek-v4-flash & 0.0 & 0.0 & 0.006 & 0.006 & 0.003 & 0.003 \\
\midrule
\multirow{5}{*}{MBPP}
  & gpt-5-mini & 0.002 & 0.228 \deltaup{} & 0.004 &  0.004 & 0.004 & 0.004 \\
  & gpt-4.1-mini & 0.005 & 0.075 \deltaup{} & 0.009 & 0.009 & 0.004 & 0.004\\
  & claude-haiku-4-5 & 0.011 & 0.081 \deltaup{} & 0.010 & 0.010 & 0.007 & 0.007\\
  & Llama-3.3-70B-Instruct & 0.012 & 0.070 \deltaup{} & 0.022 & 0.022 & 0.017 & 0.017\\
  & deepseek-v4-flash & 0.009 & 0.102 \deltaup{} & 0.005 & 0.005 & 0.004 & 0.004\\
\midrule
\multirow{5}{*}{BigCodeBench}
  & gpt-5-mini & 0.01 & 0.286 \deltaup{} & 0.007 & 0.007 & 0.008 & 0.008 \\
  & gpt-4.1-mini & 0.019 & 0.064 \deltaup{} & 0.014 & 0.014 & 0.012 & 0.012\\
  & claude-haiku-4-5 & 0.024 & 0.040 \deltaup{} & 0.007 & 0.007 & 0.005 &  0.005 \\
  & Llama-3.3-70B-Instruct & 0.030 & 0.044 \deltaup{} & 0.018 & 0.018  & 0.018 & 0.018  \\
  & deepseek-v4-flash & 0.013 & 0.074 \deltaup{} & 0.010 & 0.010 & 0.010 & 0.010\\
\midrule
\multirow{5}{*}{NaturalCodeBench}
  & gpt-5-mini & 0.0 & 0.062 \deltaup{} & 0.0 & 0.0 & 0.0 & 0.0 \\
  & gpt-4.1-mini & 0.027 & 0.00 \deltadown{} & 0.044 & 0.044 & 0.04 & 0.04\\
  & claude-haiku-4-5 & 0.0005 & 0.10 \deltaup{} & 0.008 & 0.008 & 0.007 & 0.007 \\
  & Llama-3.3-70B-Instruct & 0.080 & 0.111 \deltaup{} & 0.11 & 0.110 & 0.055 & 0.055\\
  & deepseek-v4-flash & 0.045 & 0.136 \deltaup{} & 0.082 & 0.082 & 0.055 & 0.055 \\
\bottomrule
\end{tabular*}

%% file: citations.bib
@ARTICLE{10136793,
  author={Ojdanic, Milos and Garg, Aayush and Khanfir, Ahmed and Degiovanni, Renzo and Papadakis, Mike and Le Traon, Yves},
  journal={IEEE Transactions on Software Engineering}, 
  title={Syntactic Versus Semantic Similarity of Artificial and Real Faults in Mutation Testing Studies}, 
  year={2023},
  volume={49},
  number={7},
  pages={3922-3938},
  doi={10.1109/TSE.2023.3277564}}

@article{10.1145/125489.125473,
author = {Offutt, A. Jefferson},
title = {Investigations of the software testing coupling effect},
year = {1992},
issue_date = {Jan. 1992},
publisher = {Association for Computing Machinery},
address = {New York, NY, USA},
volume = {1},
number = {1},
issn = {1049-331X},
url = {https://doi.org/10.1145/125489.125473},
doi = {10.1145/125489.125473},
journal = {ACM Trans. Softw. Eng. Methodol.},
month = jan,
pages = {5–20},
numpages = {16}
}

@inproceedings{ZhuLLZLJ024,
  author       = {Yuqi Zhu and
                  Jia Li and
                  Ge Li and
                  Yunfei Zhao and
                  Jia Li and
                  Zhi Jin and
                  Hong Mei},
  editor       = {Michael J. Wooldridge and
                  Jennifer G. Dy and
                  Sriraam Natarajan},
  title        = {Hot or Cold? Adaptive Temperature Sampling for Code Generation with
                  Large Language Models},
  booktitle    = {Thirty-Eighth {AAAI} Conference on Artificial Intelligence, {AAAI}
                  2024, Thirty-Sixth Conference on Innovative Applications of Artificial
                  Intelligence, {IAAI} 2024, Fourteenth Symposium on Educational Advances
                  in Artificial Intelligence, {EAAI} 2014, February 20-27, 2024, Vancouver,
                  Canada},
  pages        = {437--445},
  publisher    = {{AAAI} Press},
  year         = {2024},
  url          = {https://doi.org/10.1609/aaai.v38i1.27798},
  doi          = {10.1609/AAAI.V38I1.27798},
  bibsource    = {dblp computer science bibliography, https://dblp.org}
}

@article{10.1007/s10664-025-10614-4,
author = {Tambon, Florian and Moradi-Dakhel, Arghavan and Nikanjam, Amin and Khomh, Foutse and Desmarais, Michel C. and Antoniol, Giuliano},
title = {Bugs in large language models generated code: an empirical study},
year = {2025},
issue_date = {Mar 2025},
publisher = {Kluwer Academic Publishers},
address = {USA},
volume = {30},
number = {3},
issn = {1382-3256},
url = {https://doi.org/10.1007/s10664-025-10614-4},
doi = {10.1007/s10664-025-10614-4},
journal = {Empirical Softw. Engg.},
month = feb,
numpages = {48}
}

@inproceedings{DBLP:conf/sigsoft/JinSTSLSS23,
  author       = {Matthew Jin and
                  Syed Shahriar and
                  Michele Tufano and
                  Xin Shi and
                  Shuai Lu and
                  Neel Sundaresan and
                  Alexey Svyatkovskiy},
  editor       = {Satish Chandra and
                  Kelly Blincoe and
                  Paolo Tonella},
  title        = {InferFix: End-to-End Program Repair with LLMs},
  booktitle    = {Proceedings of the 31st {ACM} Joint European Software Engineering
                  Conference and Symposium on the Foundations of Software Engineering,
                  {ESEC/FSE} 2023, San Francisco, CA, USA, December 3-9, 2023},
  pages        = {1646--1656},
  publisher    = {{ACM}},
  year         = {2023},
  url          = {https://doi.org/10.1145/3611643.3613892},
  doi          = {10.1145/3611643.3613892},
  bibsource    = {dblp computer science bibliography, https://dblp.org}
}

@inproceedings{YangJWLYNP24,
  author       = {John Yang and
                  Carlos E. Jimenez and
                  Alexander Wettig and
                  Kilian Lieret and
                  Shunyu Yao and
                  Karthik Narasimhan and
                  Ofir Press},
  editor       = {Amir Globersons and
                  Lester Mackey and
                  Danielle Belgrave and
                  Angela Fan and
                  Ulrich Paquet and
                  Jakub M. Tomczak and
                  Cheng Zhang},
  title        = {SWE-agent: Agent-Computer Interfaces Enable Automated Software Engineering},
  booktitle    = {Advances in Neural Information Processing Systems 37: Annual Conference
                  on Neural Information Processing Systems 2024, NeurIPS 2024, Vancouver,
                  BC, Canada, December 10 - 15, 2024},
  year         = {2024},
  url          = {http://papers.nips.cc/paper\_files/paper/2024/hash/5a7c947568c1b1328ccc5230172e1e7c-Abstract-Conference.html},
  bibsource    = {dblp computer science bibliography, https://dblp.org}
}

@inproceedings{NEURIPS2022_d0c6bc64,
 author = {Wu, Yuhuai and Jiang, Albert Qiaochu and Li, Wenda and Rabe, Markus and Staats, Charles and Jamnik, Mateja and Szegedy, Christian},
 booktitle = {Advances in Neural Information Processing Systems},
 editor = {S. Koyejo and S. Mohamed and A. Agarwal and D. Belgrave and K. Cho and A. Oh},
 pages = {32353--32368},
 publisher = {Curran Associates, Inc.},
 title = {Autoformalization with Large Language Models},
 url = {https://proceedings.neurips.cc/paper\_files/paper/2022/file/d0c6bc641a56bebee9d985b937307367-Paper-Conference.pdf},
 volume = {35},
 year = {2022}
}

@article{DBLP:journals/pacmpl/LiHZQ24,
  author       = {Haonan Li and
                  Yu Hao and
                  Yizhuo Zhai and
                  Zhiyun Qian},
  title        = {Enhancing Static Analysis for Practical Bug Detection: An LLM-Integrated
                  Approach},
  journal      = {Proc. {ACM} Program. Lang.},
  volume       = {8},
  number       = {{OOPSLA1}},
  pages        = {474--499},
  year         = {2024},
  url          = {https://doi.org/10.1145/3649828},
  doi          = {10.1145/3649828},
  bibsource    = {dblp computer science bibliography, https://dblp.org}
}

@article{jiang2024surveylargelanguagemodels,
author = {Jiang, Juyong and Wang, Fan and Shen, Jiasi and Kim, Sungju and Kim, Sunghun},
title = {A Survey on Large Language Models for Code Generation},
year = {2026},
issue_date = {February 2026},
publisher = {Association for Computing Machinery},
address = {New York, NY, USA},
volume = {35},
number = {2},
issn = {1049-331X},
url = {https://doi.org/10.1145/3747588},
doi = {10.1145/3747588},
journal = {ACM Trans. Softw. Eng. Methodol.},
month = jan,
articleno = {58},
numpages = {72}
}

@ARTICLE{6312836,
  author={Goodenough, John B. and Gerhart, Susan L.},
  journal={IEEE Transactions on Software Engineering}, 
  title={Toward a theory of test data selection}, 
  year={1975},
  volume={SE-1},
  number={2},
  pages={156-173},
  doi={10.1109/TSE.1975.6312836}}

@misc{konstantinou2026llmbasedtestgenerationtechniques,
      title={How well LLM-based test generation techniques perform with newer LLM versions?}, 
      author={Michael Konstantinou and Renzo Degiovanni and Mike Papadakis},
      year={2026},
      eprint={2601.09695},
      archivePrefix={arXiv},
      primaryClass={cs.SE},
      url={https://arxiv.org/abs/2601.09695}, 
}

@inproceedings{someoliayi2019program,
  title={Program state coverage: A test coverage metric based on executed program states},
  author={Someoliayi, Khashayar Etemadi and Jalali, Sajad and Mahdieh, Mostafa and Mirian-Hosseinabadi, Seyed-Hassan},
  booktitle={2019 IEEE 26th International Conference on Software Analysis, Evolution and Reengineering (SANER)},
  pages={584--588},
  year={2019},
  organization={IEEE}
}

@article{molina2025state,
  title={State Field Coverage: A Metric for Oracle Quality},
  author={Molina, Facundo and Aguirre, Nazareno and Gorla, Alessandra},
  journal={arXiv preprint arXiv:2510.03071},
  year={2025}
}

@article{10.1145/267580.267590,
author = {Zhu, Hong and Hall, Patrick A. V. and May, John H. R.},
title = {Software unit test coverage and adequacy},
year = {1997},
issue_date = {Dec. 1997},
publisher = {Association for Computing Machinery},
address = {New York, NY, USA},
volume = {29},
number = {4},
issn = {0360-0300},
url = {https://doi.org/10.1145/267580.267590},
doi = {10.1145/267580.267590},
journal = {ACM Comput. Surv.},
month = dec,
pages = {366–427},
numpages = {62}
}

@book{myers2006art,
  title={The art of software testing},
  author={Myers, Glenford J},
  year={2006},
  publisher={John Wiley \& Sons}
}

@article{larbi2025prompts,
  title={When prompts go wrong: Evaluating code model robustness to ambiguous, contradictory, and incomplete task descriptions},
  author={Larbi, Maya and Akli, Amal and Papadakis, Mike and Bouyousfi, Rihab and Cordy, Maxime and Sarro, Federica and Traon, Yves Le},
  journal={arXiv preprint arXiv:2507.20439},
  year={2025}
}

@inproceedings{chekam2017empirical,
  title={An empirical study on mutation, statement and branch coverage fault revelation that avoids the unreliable clean program assumption},
  author={Chekam, Thierry Titcheu and Papadakis, Mike and Le Traon, Yves and Harman, Mark},
  booktitle={2017 IEEE/ACM 39th International Conference on Software Engineering (ICSE)},
  pages={597--608},
  year={2017},
  organization={IEEE}
}

@article{titcheu2020selecting,
  title={Selecting fault revealing mutants},
  author={Titcheu Chekam, Thierry and Papadakis, Mike and Bissyand{\'e}, Tegawend{\'e} F and Le Traon, Yves and Sen, Koushik},
  journal={Empirical Software Engineering},
  volume={25},
  number={1},
  pages={434--487},
  year={2020},
  publisher={Springer}
}

@inproceedings{ma2020commit,
  title={Commit-aware mutation testing},
  author={Ma, Wei and Laurent, Thomas and Ojdani{\'c}, Milo{\v{s}} and Chekam, Thierry Titcheu and Ventresque, Anthony and Papadakis, Mike},
  booktitle={2020 IEEE International Conference on Software Maintenance and Evolution (ICSME)},
  pages={394--405},
  year={2020},
  organization={IEEE}
}

@article{abs-2403-16218,
  author       = {Juan Altmayer Pizzorno and
                  Emery D. Berger},
  title        = {CoverUp: Coverage-Guided LLM-Based Test Generation},
  journal      = {CoRR},
  volume       = {abs/2403.16218},
  year         = {2024},
  url          = {https://doi.org/10.48550/arXiv.2403.16218},
  doi          = {10.48550/ARXIV.2403.16218},
  eprinttype    = {arXiv},
  eprint       = {2403.16218},
  bibsource    = {dblp computer science bibliography, https://dblp.org}
}

@inproceedings{sapozhnikov2024testspark,
  title={TestSpark: IntelliJ IDEA's Ultimate Test Generation Companion},
  author={Sapozhnikov, Arkadii and Olsthoorn, Mitchell and Panichella, Annibale and Kovalenko, Vladimir and Derakhshanfar, Pouria},
  booktitle={Proceedings of the 2024 IEEE/ACM 46th International Conference on Software Engineering: Companion Proceedings},
  pages={30--34},
  year={2024}
}

@inproceedings{WangL0J24,
  author       = {Zejun Wang and
                  Kaibo Liu and
                  Ge Li and
                  Zhi Jin},
  editor       = {Vladimir Filkov and
                  Baishakhi Ray and
                  Minghui Zhou},
  title        = {{HITS:} High-coverage LLM-based Unit Test Generation via Method Slicing},
  booktitle    = {Proceedings of the 39th {IEEE/ACM} International Conference on Automated
                  Software Engineering, {ASE} 2024, Sacramento, CA, USA, October 27
                  - November 1, 2024},
  pages        = {1258--1268},
  publisher    = {{ACM}},
  year         = {2024},
  url          = {https://doi.org/10.1145/3691620.3695501},
  doi          = {10.1145/3691620.3695501},
}

@article{ciupaetal,
author = {Ciupa, I. and Pretschner, A. and Oriol, M. and Leitner, A. and Meyer, B.},
title = {On the number and nature of faults found by random testing},
journal = {Software Testing, Verification and Reliability},
volume = {21},
number = {1},
pages = {3-28},
doi = {https://doi.org/10.1002/stvr.415},
url = {https://onlinelibrary.wiley.com/doi/abs/10.1002/stvr.415},
eprint = {https://onlinelibrary.wiley.com/doi/pdf/10.1002/stvr.415},
year = {2011}
}

@Inbook{Wei2012,
author="Wei, Yi
and Meyer, Bertrand
and Oriol, Manuel",
editor="Meyer, Bertrand
and Nordio, Martin",
title="Is Branch Coverage a Good Measure of Testing Effectiveness?",
bookTitle="Empirical Software Engineering and Verification: International Summer Schools, LASER 2008-2010, Elba Island, Italy, Revised Tutorial Lectures",
year="2012",
publisher="Springer Berlin Heidelberg",
address="Berlin, Heidelberg",
pages="194--212",
isbn="978-3-642-25231-0",
doi="10.1007/978-3-642-25231-0\_5",
url="https://doi.org/10.1007/978-3-642-25231-0\_5"
}

@INPROCEEDINGS{6606563,
  author={Hassan, Mohammad Mahdi and Andrews, James H.},
  booktitle={2013 35th International Conference on Software Engineering (ICSE)}, 
  title={Comparing Multi-Point Stride Coverage and dataflow coverage}, 
  year={2013},
  volume={},
  number={},
  pages={172-181},
  doi={10.1109/ICSE.2013.6606563}}

@article{ANAND20131978,
title = {An orchestrated survey of methodologies for automated software test case generation},
journal = {Journal of Systems and Software},
volume = {86},
number = {8},
pages = {1978-2001},
year = {2013},
issn = {0164-1212},
doi = {https://doi.org/10.1016/j.jss.2013.02.061},
url = {https://www.sciencedirect.com/science/article/pii/S0164121213000563},
author = {Saswat Anand and Edmund K. Burke and Tsong Yueh Chen and John Clark and Myra B. Cohen and Wolfgang Grieskamp and Mark Harman and Mary Jean Harrold and Phil McMinn and Antonia Bertolino and J. {Jenny Li} and Hong Zhu}
}

@inproceedings{10.1145/2568225.2568278,
author = {Gopinath, Rahul and Jensen, Carlos and Groce, Alex},
title = {Code coverage for suite evaluation by developers},
year = {2014},
isbn = {9781450327565},
publisher = {Association for Computing Machinery},
address = {New York, NY, USA},
url = {https://doi.org/10.1145/2568225.2568278},
doi = {10.1145/2568225.2568278},
booktitle = {Proceedings of the 36th International Conference on Software Engineering},
pages = {72–82},
numpages = {11},
location = {Hyderabad, India},
series = {ICSE 2014}
}

@inproceedings{10.1145/2568225.2568271,
author = {Inozemtseva, Laura and Holmes, Reid},
title = {Coverage is not strongly correlated with test suite effectiveness},
year = {2014},
isbn = {9781450327565},
publisher = {Association for Computing Machinery},
address = {New York, NY, USA},
url = {https://doi.org/10.1145/2568225.2568271},
doi = {10.1145/2568225.2568271},
booktitle = {Proceedings of the 36th International Conference on Software Engineering},
pages = {435–445},
numpages = {11},
location = {Hyderabad, India},
series = {ICSE 2014}
}

@inproceedings{10.1145/120807.120821,
author = {Frankl, Phyllis G. and Weiss, Stewart N.},
title = {An experimental comparison of the effectiveness of the all-uses and all-edges adequacy criteria},
year = {1991},
isbn = {089791449X},
publisher = {Association for Computing Machinery},
address = {New York, NY, USA},
url = {https://doi.org/10.1145/120807.120821},
doi = {10.1145/120807.120821},
booktitle = {Proceedings of the Symposium on Testing, Analysis, and Verification},
pages = {154–164},
numpages = {11},
location = {Victoria, British Columbia, Canada},
series = {TAV4}
}

@article{FRANKL1997235,
title = {All-uses vs mutation testing: An experimental comparison of effectiveness},
journal = {Journal of Systems and Software},
volume = {38},
number = {3},
pages = {235-253},
year = {1997},
issn = {0164-1212},
doi = {https://doi.org/10.1016/S0164-1212(96)00154-9},
url = {https://www.sciencedirect.com/science/article/pii/S0164121296001549},
author = {Phyllis G. Frankl and Stewart N. Weiss and Cang Hu}
}

@inproceedings{10.1145/288195.288298,
author = {Frankl, Phyllis G. and Iakounenko, Oleg},
title = {Further empirical studies of test effectiveness},
year = {1998},
isbn = {1581131089},
publisher = {Association for Computing Machinery},
address = {New York, NY, USA},
url = {https://doi.org/10.1145/288195.288298},
doi = {10.1145/288195.288298},
booktitle = {Proceedings of the 6th ACM SIGSOFT International Symposium on Foundations of Software Engineering},
pages = {153–162},
numpages = {10},
location = {Lake Buena Vista, Florida, USA},
series = {SIGSOFT '98/FSE-6}
}

@incollection{PAPADAKIS2019275,
title = {Chapter Six - Mutation Testing Advances: An Analysis and Survey},
editor = {Atif M. Memon},
series = {Advances in Computers},
publisher = {Elsevier},
volume = {112},
pages = {275-378},
year = {2019},
issn = {0065-2458},
doi = {https://doi.org/10.1016/bs.adcom.2018.03.015},
url = {https://www.sciencedirect.com/science/article/pii/S0065245818300305},
author = {Mike Papadakis and Marinos Kintis and Jie Zhang and Yue Jia and Yves Le Traon and Mark Harman}
}

@book{ammann2017introduction,
  title={Introduction to software testing},
  author={Ammann, Paul and Offutt, Jeff},
  year={2017},
  publisher={Cambridge University Press}
}

@inproceedings{10.1145/2635868.2635929,
author = {Just, Ren\'{e} and Jalali, Darioush and Inozemtseva, Laura and Ernst, Michael D. and Holmes, Reid and Fraser, Gordon},
title = {Are mutants a valid substitute for real faults in software testing?},
year = {2014},
isbn = {9781450330565},
publisher = {Association for Computing Machinery},
address = {New York, NY, USA},
url = {https://doi.org/10.1145/2635868.2635929},
doi = {10.1145/2635868.2635929},
booktitle = {Proceedings of the 22nd ACM SIGSOFT International Symposium on Foundations of Software Engineering},
pages = {654–665},
numpages = {12},
location = {Hong Kong, China},
series = {FSE 2014}
}

@article{dakhel2024effective,
  title={Effective test generation using pre-trained large language models and mutation testing},
  author={Dakhel, Arghavan Moradi and Nikanjam, Amin and Majdinasab, Vahid and Khomh, Foutse and Desmarais, Michel C},
  journal={Information and Software Technology},
  volume={171},
  pages={107468},
  year={2024},
  publisher={Elsevier}
}

@article{vathana2026llm,
  title={LLM vs. Human Unit Tests: Fault Detection on Real Python Bugs},
  author={Vathana, Phouvadeth and Bhatt, Prapti and Patel, Rishi and Eisty, Nasir U},
  journal={arXiv preprint arXiv:2606.08588},
  year={2026}
}

@article{doi.org/10.1002/smr.70034,
author = {Liu, Xiangyue and Sun, Xiaobing and Bo, Lili and Hu, Yufei and Liu, Xinwei and Ye, Zhenlei},
title = {Evaluating the Test Adequacy of Benchmarks for LLMs on Code Generation},
journal = {Journal of Software: Evolution and Process},
volume = {37},
number = {7},
pages = {e70034},
doi = {https://doi.org/10.1002/smr.70034},
url = {https://onlinelibrary.wiley.com/doi/abs/10.1002/smr.70034},
eprint = {https://onlinelibrary.wiley.com/doi/pdf/10.1002/smr.70034},
year = {2025}
}

@article{reinhart2025llms,
  title={Do LLMs write like humans? Variation in grammatical and rhetorical styles},
  author={Reinhart, Alex and Markey, Ben and Laudenbach, Michael and Pantusen, Kachatad and Yurko, Ronald and Weinberg, Gordon and Brown, David West},
  journal={Proceedings of the National Academy of Sciences},
  volume={122},
  number={8},
  pages={e2422455122},
  year={2025},
  publisher={National Academy of Sciences}
}

@inproceedings{zhuo2025bigcodebench,
  title={Bigcodebench: Benchmarking code generation with diverse function calls and complex instructions},
  author={Zhuo, Terry Yue and Vu, Minh Chien and Chim, Jenny and Hu, Han and Yu, Wenhao and Widyasari, Ratnadira and Yusuf, Imam Nur Bani and Zhan, Haolan and He, Junda and Paul, Indraneil and others},
  booktitle={International Conference on Learning Representations},
  volume={2025},
  pages={66602--66656},
  year={2025}
}

@article{liu2024your,
  title={Is your code generated by chatgpt really correct? rigorous evaluation of large language models for code generation},
  author={Liu, Jiawei and Xia, Chunqiu Steven and Wang, Yuyao and Zhang, Lingming},
  journal={Advances in Neural Information Processing Systems},
  volume={36},
  year={2024}
}

@article{chen2021evaluating,
  title={Evaluating large language models trained on code},
  author={Chen, Mark and Tworek, Jerry and Jun, Heewoo and Yuan, Qiming and Pinto, Henrique Ponde De Oliveira and Kaplan, Jared and Edwards, Harri and Burda, Yuri and Joseph, Nicholas and Brockman, Greg and others},
  journal={arXiv preprint arXiv:2107.03374},
  year={2021}
}

@article{austin2021program,
  title={Program synthesis with large language models},
  author={Austin, Jacob and Odena, Augustus and Nye, Maxwell and Bosma, Maarten and Michalewski, Henryk and Dohan, David and Jiang, Ellen and Cai, Carrie and Terry, Michael and Le, Quoc and others},
  journal={arXiv preprint arXiv:2108.07732},
  year={2021}
}

@article{zhang2024naturalcodebench,
  title={Naturalcodebench: Examining coding performance mismatch on humaneval and natural user prompts},
  author={Zhang, Shudan and Zhao, Hanlin and Liu, Xiao and Zheng, Qinkai and Qi, Zehan and Gu, Xiaotao and Zhang, Xiaohan and Dong, Yuxiao and Tang, Jie},
  journal={arXiv preprint arXiv:2405.04520},
  year={2024}
}

@misc{konstantinou2025yateroletestrepair,
      title={YATE: The Role of Test Repair in LLM-Based Unit Test Generation}, 
      author={Michael Konstantinou and Renzo Degiovanni and Jie M. Zhang and Mark Harman and Mike Papadakis},
      year={2025},
      eprint={2507.18316},
      archivePrefix={arXiv},
      primaryClass={cs.SE},
      url={https://arxiv.org/abs/2507.18316}, 
}

@misc{konstantinou2024llmsgeneratetestoracles,
      title={Do LLMs generate test oracles that capture the actual or the expected program behaviour?}, 
      author={Michael Konstantinou and Renzo Degiovanni and Mike Papadakis},
      year={2024},
      eprint={2410.21136},
      archivePrefix={arXiv},
      primaryClass={cs.SE},
      url={https://arxiv.org/abs/2410.21136}, 
}

@article{huang2024measuring,
  title={Measuring the influence of incorrect code on test generation},
  author={Huang, Dong and Zhang, Jie M and Harman, Mark and Du, Mingzhe and Cui, Heming},
  journal={arXiv preprint arXiv:2409.09464},
  year={2024}
}

@inproceedings{papadakis2015trivial,
  title={Trivial compiler equivalence: A large scale empirical study of a simple, fast and effective equivalent mutant detection technique},
  author={Papadakis, Mike and Jia, Yue and Harman, Mark and Le Traon, Yves},
  booktitle={2015 IEEE/ACM 37th IEEE International Conference on Software Engineering},
  volume={1},
  pages={936--946},
  year={2015},
  organization={IEEE}
}

@inproceedings{li2009experimental,
  title={An experimental comparison of four unit test criteria: Mutation, edge-pair, all-uses and prime path coverage},
  author={Li, Nan and Praphamontripong, Upsorn and Offutt, Jeff},
  booktitle={2009 International Conference on Software Testing, Verification, and Validation Workshops},
  pages={220--229},
  year={2009},
  organization={IEEE}
}

@article{chang2026test,
  title={Test vs Mutant: Adversarial LLM Agents for Robust Unit Test Generation},
  author={Chang, Pengyu and Fang, Yixiong and Chen, Silin and Shi, Yuling and Shen, Beijun and Gu, Xiaodong},
  journal={arXiv preprint arXiv:2602.08146},
  year={2026}
}

@article{akli2026defective,
  title={Defective Task Descriptions in LLM-Based Code Generation: Detection and Analysis},
  author={Akli, Amal and Papadakis, Mike and Cordy, Maxime and Traon, Yves Le},
  journal={arXiv preprint arXiv:2604.24703},
  year={2026}
}

@article{DoER05,
  author       = {Hyunsook Do and
                  Sebastian G. Elbaum and
                  Gregg Rothermel},
  title        = {Supporting Controlled Experimentation with Testing Techniques: An
                  Infrastructure and its Potential Impact},
  journal      = {Empir. Softw. Eng.},
  volume       = {10},
  number       = {4},
  pages        = {405--435},
  year         = {2005},
  url          = {https://doi.org/10.1007/s10664-005-3861-2},
  doi          = {10.1007/S10664-005-3861-2},
}

@article{ChekamPBTS20,
  author       = {Thierry Titcheu Chekam and
                  Mike Papadakis and
                  Tegawend{\'{e}} F. Bissyand{\'{e}} and
                  Yves Le Traon and
                  Koushik Sen},
  title        = {Selecting fault revealing mutants},
  journal      = {Empir. Softw. Eng.},
  volume       = {25},
  number       = {1},
  pages        = {434--487},
  year         = {2020},
  url          = {https://doi.org/10.1007/s10664-019-09778-7},
  doi          = {10.1007/S10664-019-09778-7},
}
